\documentclass[aps,twocolumn,superscriptaddress,amsmath,showpacs,tightenlines]{revtex4}
\usepackage{graphicx}
\usepackage{epsfig}

\begin{document}
%
\title{Efficient purification protocols using iSWAP gates in solid-state qubits}

\author{Tetsufumi Tanamoto}
\affiliation{Corporate R \& D center, Toshiba Corporation,
Saiwai-ku, Kawasaki 212-8582, Japan}

\author{Koji Maruyama}
\affiliation{Advanced Science Institute, The Institute of Physical
and Chemical Research (RIKEN), Wako-shi, Saitama 351-0198,
Japan}

\author{Yu-xi Liu}
\affiliation{Advanced Science Institute, The Institute of Physical
and Chemical Research (RIKEN), Wako-shi, Saitama 351-0198,
Japan}
\affiliation{CREST, Japan Science and Technology Agency (JST), 
Kawaguchi, Saitama 332-0012, Japan}

\author{Xuedong Hu}
\affiliation{Department of Physics, University at Buffalo, SUNY,
Buffalo, New York 14260-1500,USA}

\author{Franco Nori}
\affiliation{Advanced Science Institute, The Institute of Physical
and Chemical Research (RIKEN), Wako-shi, Saitama 351-0198,
Japan}
\affiliation{CREST, Japan Science and Technology Agency (JST), 
Kawaguchi, Saitama 332-0012, Japan}
\affiliation{Physics Department, Center for Theoretical Physics, 
Center for the Study of Complex Systems, The University of
Michigan, Ann Arbor, Michigan 48109-1040, USA}

\date{\today}

\begin{abstract}
We show an efficient purification protocol in solid-state 
qubits by replacing the usual bilateral CNOT gate by 
the bilateral iSWAP gate. We also show that this replacement 
can be applied to breeding and hashing protocols, which are 
useful for quantum state purification. 
These replacements reduce the number of 
fragile and cumbersome two-qubit operations, making more feasible 
quantum-information-processing with solid-state qubits.
As examples, we also present quantitative analyses for the required time 
to perform state purification using either superconducting or 
semiconducting qubits.
\end{abstract}
\pacs{03.67.Lx, 03.67.Mn, 73.21.La}
\maketitle

\section{Introduction}
%
%
Quantum communications, such as quantum 
teleportation~\cite{Bennett2} and secure quantum cryptography~\cite{Ekert},
between two parties (Alice and Bob), require that qubits in highly entangled states, 
such as Bell states, be shared between the parties.  
The entanglement purification protocols  
proposed by Bennett {\it et al.}~\cite{Bennett} and Deutsch 
{\it et al.}~\cite{Deutsch} are therefore not only important contributions
to the theory of quantum information, but also essential ingredients to 
applications such as quantum communications.  Starting from partially 
entangled states, these protocols distill near-maximally entangled states 
shared by distant parties.  More specifically, in such a purification protocol, multiple 
pairs of qubits in impure entangled states are initially supplied, from which purified 
pairs are then obtained after sacrificing some of the impure pairs.

In each step of an entanglement purification protocol, local quantum computers 
have to carry out several single-qubit rotations and two-qubit operations 
on the local qubits at the sites of Alice and Bob, respectively.  In particular,
controlled-NOT (CNOT) gates play a major role in these purification protocols 
(as well as in other fields of quantum information and computation).  
In purification protocols \cite{Bennett,Deutsch}, Alice and Bob 
repeat a process in which, after choosing two shared entangled
pairs in mixed states, they 
bilaterally  apply CNOT gates to their two local qubits that
 belong to the shared pairs, and measure one of the pairs.  
If the measured qubits are in either 
the $|00\rangle$ or $|11\rangle$ state, then the unmeasured pair is 
forwarded to the next step; otherwise the unmeasured pair is discarded.  
In the more efficient Deutsch protocol~\cite{Deutsch}, tens of such repetitions 
are needed, which means that a corresponding number of CNOT gates needs 
to be employed, and they should work with very low error rate. 
%
%

For most solid-state qubits, two-qubit interactions are quite delicate and are 
difficult to control without error and decoherence.  
%
%
As such, creating quantum algorithms that employ {\it fewer} two-qubit operations is 
important to the successful construction of solid state quantum information 
processors.  This optimization of the algorithmic aspects demands a closer inspection 
of the omnipresent CNOT gate, a standard two-qubit gate. The CNOT gate is most 
conveniently generated from Ising-interactions.  However, general solid-state
interqubit interactions are not of the Ising-type.  Instead, they are often in the 
form of the Heisenberg exchange ( e.g., as in electrically tuned quantum dots) or 
XY model ( e.g., cavity-coupled semiconducting quantum dots (QD)~\cite{Imamoglu} or  
superconducting Josephson qubits~\cite{You}).  In general, 
when a CNOT gate is constructed using the Heisenberg exchange or the XY interaction, 
{\it at least twice} the number of two-qubit interactions have to be invoked with 
complicated pulse sequences.  A key question is thus whether 
it is possible to devise quantum 
algorithms that take better advantage of these two particular qubit interactions, instead 
of relying exclusively on the standard but cumbersome CNOT gate.
 
A further incentive to study XY-model-based quantum algorithms lies in the 
recent advances in cavity-coupling of Josephson superconducting 
qubits (see, e.g.,~\cite{You,Houck,NTT,You2}) and cavity quantum electrodynamics 
(QED) of Josephson qubits, because it is relatively easy to 
reach the strongly interacting regime for these systems.  Since cavity QED plays 
an important role for information exchange between static and flying qubits 
in quantum communication networks, and the effective interaction between 
cavity-coupled qubits is described by the XY-model, the development of XY-model-based
quantum algorithms would pave the way for an easier integration of solid state
qubits into a quantum communication network.

In this paper we study {\it how to build entanglement purification protocols
based on a two-qubit gate that can be easily generated by the XY interaction}. 
%
%
It is important to note that it is relatively easy to generate the iSWAP gate in the XY model. 
Indeed, the CNOT gate is built using {\it two} iSWAP gates {\it and} several single-qubit
gates \cite{Schuch}.  Here we show that the bilateral CNOT gate (BCNOT) 
used in entanglement purification protocols
can be replaced by a bilateral iSWAP gate (BiSWAP). 
%
%
For solid-state qubits with XY inter-qubit interactions, this change of 
gates leads to {\it a significant simplification of each step of the entanglement
purification protocol, and to a much higher robustness of the protocol.}
%
%
Furthermore, purification protocols are often followed by hashing or
breeding protocols.  Here we show that 
the bilateral CNOT gates in the hashing or breeding protocol can also
be replaced by bilateral iSWAP gates. 
In addition, we also discuss a purification protocol using 
$\sqrt{\rm SWAP}$ gates, which are the basic typical operations for qubits
that are coupled via Heisenberg exchange interactions.

The rest of the paper is organized as follows.
In section II, we formulate the iSWAP gate from the XY-model Hamiltonian and 
the $\sqrt{\rm SWAP}$ gate in the Heisenberg model.
In section III, we show a purification protocol based on the iSWAP gate and 
discuss the effect of gate errors.
In section~\ref{sec-hashing}, we discuss the replacement of CNOT gates 
by iSWAP gates in the hashing and breeding protocols.
In section~\ref{sec-Bell}, we show an effective method of generating 
the four Bell states based on the iSWAP gate. 
%
%
In section~\ref{sec-example}, we give four examples of the 
application of the present method.
%
%
Section~\ref{sec-discussion} and \ref{sec-summary} present discussions and a 
summary.
In the appendix, we summarize the derivation of the XY interaction 
from a general qubit-cavity Hamiltonian.
Let us note that we do not assume any particular method 
for distributing entangled qubits. 
In the following discussions, noisy entanglement is taken as a resource.

\begin{figure}
\includegraphics[width=6.0cm]{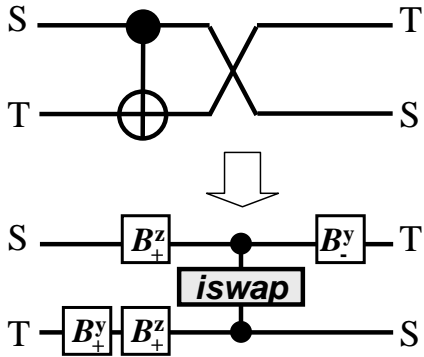}
\caption{Replacement of a bilateral CNOT (BCNOT) gate by a bilateral iSWAP (BiSWAP) gate. 
This figure shows the protocol for one of the parties. The complete protocol 
is achieved by the execution of the same operation at both ends.
Here, we define 
bilateral single-qubit $\pm \pi/2$ rotations for Alice and Bob about the $x$, $y$ 
and $z$ axes as those in Ref.~\cite{Bennett}, denoted by $B^x_\pm$, $B^y_\pm$ 
and $B^z_\pm$, respectively.
} 
\label{replace}
\end{figure}

\section{iSWAP gate in the XY model and the $\sqrt{\rm SWAP}$ gate 
in the Heisenberg model} 
\label{sec-formulation}

In this Section we formulate the iSWAP gate from the XY model, 
and estimate the time required to obtain a conventional CNOT gate using an iSWAP gate. 
We also consider the case of the $\sqrt{\rm SWAP}$ gate from 
the Heisenberg model.

The Hamiltonian of a coupled qubit-cavity system is typically given 
by the Jaynes-Cummings Hamiltonian, representing a linear 
interaction between a two-level system and a bosonic degree of 
freedom for the cavity, such as photons.  When two qubits are
coupled to the same cavity mode, the effective two-qubit 
interaction is described by the XY model.
A derivation of the XY interaction from the 
Jaynes-Cummings Hamiltonian is 
given in the Appendix. 

The XY model is expressed by the Hamiltonian
$
H_{xy}=\sum_{i<j}H_{xy}^{(ij)}
$
with 
\begin{equation}
H_{xy}^{(ij)}= J_{ij}(
\sigma_{i}^x\sigma_{j}^x+
\sigma_{i}^y\sigma_{j}^y),
\end{equation}
where $\sigma_i^\alpha$ $(\alpha=x,y,z)$ are the 
Pauli matrices acting on the $i$-th qubit with 
basis $|0\rangle =|\downarrow\rangle$ and 
$|1\rangle =|\uparrow\rangle$. 
Two-qubit operations produced by $H_{xy}^{(12)}$ acting on 
qubits `1' and `2' can thus be 
expressed as
\begin{equation}
U_{xy}^{(12)}(t)=e^{itH_{xy}^{(12)}}
=\left(
\begin{array}{cccc}
1 & 0 & 0 & 0 \\
0 & \cos 2Jt & i\sin 2Jt& 0 \\
0 & i\sin 2Jt  & \cos 2Jt & 0 \\
0 & 0 & 0 & 1 
\end{array}
\right)
\label{iswap}
\end{equation}
with 
$J=J_{12}$.
Note that the iSWAP gate is obtained when 
$t=\tau_{\rm iswap}\equiv \pi/(4J)$ 
such that
\begin{eqnarray}  
|00\rangle \rightarrow |00\rangle, \  
|11\rangle \rightarrow |11\rangle, \nonumber \\
|01\rangle \rightarrow i|10\rangle, \  
|10\rangle \rightarrow i|01\rangle.
\end{eqnarray}
The conventional CNOT gate is constructed with two iSWAP gates and four single-qubit
rotations: 
\begin{equation}
U_{\rm cnot}^{(12)}=e^{-i\frac{\pi}{4}\sigma_{1}^z}
e^{i\frac{\pi}{4}\sigma_{2}^x}
e^{i\frac{\pi}{4}\sigma_{2}^z} 
U_{\rm iswap} e^{-i\frac{\pi}{4}\sigma_{1}^x} U_{\rm iswap} 
e^{i\frac{\pi}{4}\sigma_{2}^z}, 
\label{cnot}
\end{equation}
with~\cite{Schuch} 
\begin{equation}
U_{\rm iswap}\equiv U_{xy}^{(12)}(t=\tau_{\rm iswap}).
\end{equation}  
Thus, in order to produce a single CNOT operation, 
we have to precisely control two two-qubit operations 
and four single-qubit rotations.
If we denote a single-qubit frequency as $\omega_{\rm rot}$, 
the time for a single-qubit rotation is typically
$\tau_{\rm rot}=\pi/(4\omega_{\rm rot})$.  The time to implement a CNOT
gate is thus:
\begin{equation}
\tau_{\rm cnot}\approx 4\tau_{\rm rot} + 2\tau_{\rm iswap}=
 \left(\frac{1}{\omega_{\rm rot}}+ \frac{2}{J}\right)\pi . 
\label{time_cnot}
\end{equation}
In this paper, we also study qubits 
whose interaction Hamiltonian is an isotropic Heisenberg form, 
written as $H_{H}=\sum_{i<j} H_H^{(ij)}$ with
\begin{equation}
H_{H}^{(ij)}= J_H(
\sigma_{i}^x\sigma_{j}^x+
\sigma_{i}^y\sigma_{j}^y
+\sigma_{i}^z\sigma_{j}^z).
\end{equation}
In this case, the typical two-qubit gate operation 
is $\sqrt{\rm SWAP}$, which
is defined by~\cite{Burkard}
\begin{equation}
U_{\sqrt{\rm swap}}\equiv U_{H}^{(12)}(t=\tau_{\sqrt{\rm swap}}).
\end{equation}
The CNOT gate is expressed by
$U_{\rm cnot}^{(12)}=
e^{-i\frac{\pi}{4}\sigma_{2}^y}
U_{\rm cpf}^{(12)}
e^{i\frac{\pi}{4}\sigma_{2}^y},
$
where the controlled phase flip (CPF) gate 
is obtained by 
\begin{equation}
U_{\rm cpf}^{(12)}=e^{-i\frac{\pi}{2}}
e^{i\frac{\pi}{4}\sigma_{1}^z} 
e^{-i\frac{\pi}{4}\sigma_{2}^z} 
U_{\sqrt{\rm swap}}
e^{-i\frac{\pi}{2}\sigma_{1}^z}
 U_{\sqrt{\rm swap}}.
\end{equation}
Thus, the time to implement a CNOT gate becomes
\begin{equation}
\tau_{\rm cnot}^{(H)}\approx 3\tau_{\rm rot}+2\tau_{\sqrt{\rm swap}}.
\end{equation}

\section{Simplification of the purification protocol}
\subsection{State purification using iSWAP gates}

In this Section we show that the two purification protocols proposed by 
Bennett {\it et al}.~\cite {Bennett} and Deutsch 
{\it et al}.~\cite{Deutsch} can be recasted using the iSWAP gate 
{\it instead of} the CNOT gate.  The initially supplied entangled pairs of qubits are 
assumed to be in a mixed state $\rho$.  The purification protocol 
proceeds recursively by choosing two entangled pairs,  
applying a bilateral CNOT gate and measuring one of the pairs (called target qubits). 
The application of the bilateral CNOT to two pairs, 
$\rho_S$ (source pair) and $\rho_T$ (target pair), is described by
\begin{equation}
\rho_S \otimes \rho_T \rightarrow U_{\rm bcnot} (\rho_S \otimes \rho_T) 
U_{\rm bcnot}^\dagger,
\end{equation}
where $U_{\rm bcnot}$ indicates that Alice and Bob bilaterally operate 
the CNOT gate on their local qubits that belong to $\rho_S$ and $\rho_T$. 
Here we use the four Bell basis states
\begin{equation}
\Phi^\pm =(|\uparrow\uparrow\rangle\pm |\downarrow\downarrow\rangle)/\sqrt{2}, \ \ \
\Psi^\pm =(|\uparrow\downarrow\rangle\pm |\downarrow\uparrow\rangle)/\sqrt{2}.
\end{equation}
Then, as an example, the bilateral CNOT works like this
\begin{equation}
U_{\rm bcnot} \Psi^+_S \Phi^-_T = \Psi^-_S \Psi^-_T
\end{equation} 
between a source pair 
\begin{equation}
\Psi^+_S =(|\uparrow_S^A\downarrow_S^B\rangle 
+|\downarrow_S^A\uparrow_S^B\rangle)/\sqrt{2},
\end{equation}
and a target pair
\begin{equation}
\Phi^-_T =(|\uparrow_T^A\uparrow_T^B\rangle 
-|\downarrow_T^A\downarrow_T^B\rangle)/\sqrt{2},
\end{equation} 
where 
$|\uparrow_\eta^A \rangle$ and 
$|\downarrow_\eta^A \rangle$
denote qubits that belong to Alice, and 
 $|\downarrow_\eta^B \rangle$ and
$|\downarrow_\eta^B \rangle$ indicate those that belong to Bob 
($\eta=S,T$). 
\begin{figure}
\includegraphics[width=7cm]{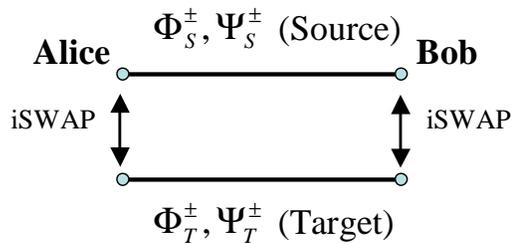}
\caption{Bilateral iSWAP (BiSWAP) gate. The iSWAP gates are 
bilaterally applied by Alice and Bob.} 
\label{Alice}
\end{figure}
%

Below we show that the conventional bilateral CNOT gate (BCNOT) 
can be replaced by the bilateral iSWAP gate (BiSWAP) together with 
a few single-qubit rotations (see Fig.~\ref{Alice}). 
First we introduce the gates involved.
The BiSWAP gate is defined as an application of the iSWAP gate at 
both locations to a pair of entangled qubits depicted in Fig.~1 
(we call one pair `source' and the other `target', as in 
Refs.~\cite{Bennett,Deutsch}).  Generally, the iSWAP gate can be 
expressed as
\begin{eqnarray}
U_{\rm iswap}&=& 
 |\uparrow_S \uparrow_T \rangle \langle \uparrow_S \uparrow_T|
+|\downarrow_S \downarrow_T \rangle \langle \downarrow_S \downarrow_T|
\nonumber \\
&+&i|\uparrow_S \downarrow_T \rangle \langle \downarrow_S \uparrow_T|
+i|\downarrow_S \uparrow_T \rangle \langle \uparrow_S \downarrow_T|\,. 
\end{eqnarray}
Here is an example of the BiSWAP gate: 
\begin{eqnarray}
U_{\rm biswap}\Phi^-_S\Phi^\pm_T  &=& 
      |\uparrow_S   \uparrow_S  \rangle |\uparrow_T \uparrow_T \rangle
  \mp |\downarrow_S \downarrow_S\rangle |\uparrow_T \uparrow_T \rangle  
\nonumber \\
   &+&|\uparrow_S   \uparrow_S  \rangle |\downarrow_T \downarrow_T \rangle 
  \mp |\downarrow_S \downarrow_S\rangle |\downarrow_T \downarrow_T \rangle 
  \nonumber \\
  &=& \Phi^{\mp}_S\Phi^+_T.
\end{eqnarray}
Similar to the BiSWAP gate, we also define 
bilateral single-qubit $\pm \pi/2$ rotations for Alice and Bob about the $x$, $y$ 
and $z$ axes as those in Ref.~\cite{Bennett}, denoted by $B^x_\pm$, $B^y_\pm$ 
and $B^z_\pm$, respectively.  For example~(the complete logic table is given in 
Table~\ref{tab:rot}), 
\begin{equation}
B_{S+}^x \Psi_S^+  =e^{i\pi\sigma_A^x /4}e^{i\pi\sigma_B^x /4}
(|\downarrow^A_S\uparrow^B_S \rangle +|\uparrow^A_S\downarrow^B_S \rangle)
=i\Phi_S^+ . 
\end{equation}
\begin{table}[h]
\caption{Bilateral rotations. Note that, 
besides the coefficient, $\pm i$, the singlet state $\Psi^-$
is unchanged. While, for the $B^x_\pm$ mapping, $\Phi^+ \leftrightarrow \Psi^+$
are exchanged. For the $B^y_\pm$ mapping,  the states $\Phi^- \leftrightarrow \Psi^+$ 
are exchanged. Finally for $B^z_\pm$ mapping, the states $\Phi^+ \leftrightarrow \Phi^-$
are exchanged.}
\begin{center}
\begin{tabular}{c|rrrr}
\hline\hline
          &$\Phi^+$      &     $\Phi^-$ &      $\Psi^+$ &      $\Psi^-$ \\ \hline
\ $B^x_{\pm}$ \ \ &\ $\pm i\Psi^+$&\    $\Phi^-$ &\ $\pm i \Phi^+$ &\      $\Psi^-$ \\
\ $B^y_{\pm}$ \ \ &\     $\Phi^+$ &\ $\mp\Psi^+$ &\ $\pm   \Phi^-$ &\      $\Psi^-$ \\
\ $B^z_{\pm}$ \ \ &\     $\Phi^-$ &\    $\Phi^+$ &\ $\mp i \Psi^+$ &\ $\mp i \Psi^-$\\
\hline\hline
\end{tabular}
\end{center}
\label{tab:rot}
\end{table}
%

\begin{table}[h]
\caption{Replacing a BCNOT by a BiSWAP.
Note that the initial state on the leftmost column is subject to
four operations or steps described in the remaining four columns}
\begin{tabular}{r|r|r|r|r}
\hline\hline
initial &    step (i) &           step (ii)   & step (iii) & final step (iv)      \\ 
 state \ & $B_{T+}^y$  & $B_{S+}^z$ $B_{T+}^z$ & BiSWAP    &  $B^y_{S-}$\\ \hline
$\Phi^+_S\Phi^+_T$& $\Phi^+_S\Phi^+_T$ & $\Phi^-_S\Phi^-_T$ & $\Phi^+_S\Phi^+_T$ & $\Phi^+_S\Phi^+_T$ \\ 
        $\Phi^-_T$&$(-1)$\ \ \ $\Psi^+_T$ & $(+i)\ \ \ \Psi^+_T$&$-\Psi^+_S\Phi^-_T$ &$\Phi^-_S\Phi^-_T$ \\ 
        $\Psi^+_T$&         $\Phi^-_T$ &         $\Phi^+_T$ & $\Phi^-_S\Phi^+_T$ & $\Psi^+_S\Phi^+_T$ \\ 
        $\Psi^-_T$&         $\Psi^-_T$ & $(-i)\ \ \  \Psi^-_T$& $\Psi^-_S\Phi^-_T$ &$\Psi^-_S\Phi^-_T$ \\ 
$\Phi^-_S\Phi^+_T$& $\Phi^-_S\Phi^+_T$ & $\Phi^+_S\Phi^-_T$ & $\Phi^+_S\Phi^-_T$ & $\Phi^+_S\Phi^-_T$ \\ 
        $\Phi^-_T$&$(-1)$\ \ \ $\Psi^+_T$ & $(+i)\ \ \ \Psi^+_T$&$-\Psi^+_S\Phi^+_T$ &$\Phi^-_S\Phi^+_T$ \\ 
        $\Psi^+_T$&         $\Phi^-_T$ &         $\Phi^+_T$ & $\Phi^-_S\Phi^-_T$ & $\Psi^+_S\Phi^-_T$ \\ 
        $\Psi^-_T$&         $\Psi^-_T$ & $(-i)\ \ \ \Psi^-_T$& $\Psi^-_S\Phi^+_T$ &$\Psi^-_S\Phi^+_T$ \\ \hline
$\Psi^+_S\Phi^+_T$& $\Psi^+_S\Phi^+_T$ & $-i\Psi^+_S\Phi^-_T$&$ \Phi^-_S\Psi^+_T$ &$\Psi^+_S\Psi^+_T$ \\ 
        $\Phi^-_T$&$(-1)$\ \ \ $\Psi^+_T$ &     $\Psi^+_T$   &$ \Psi^-_S\Psi^-_T$ & $\Psi^-_S\Psi^-_T$ \\ 
        $\Psi^+_T$&         $\Phi^-_T$ & $(-i) \ \ \ \Phi^+_T$ &$ \Phi^+_S\Psi^+_T$ & $\Phi^+_S\Psi^+_T$ \\ 
        $\Psi^-_T$&         $\Psi^-_T$ & $(-1)  \ \ \ \Psi^-_T$ &$-\Psi^+_S\Psi^-_T$ & $\Phi^-_S\Psi^-_T$ \\ 
$\Psi^-_S\Phi^+_T$& $\Psi^-_S\Phi^+_T$ & $-i\Psi^-_S\Phi^-_T$&$ \Phi^-_S\Psi^-_T$ &$\Psi^+_S\Psi^-_T$ \\ 
        $\Phi^-_T$&$(-1)$\ \ \ $\Psi^+_T$ &    $ \Psi^+_T$        &$ \Psi^-_S\Psi^+_T$ & $\Psi^-_S\Psi^+_T$ \\ 
        $\Psi^+_T$&         $\Phi^-_T$ & $(-i)\ \ \ \Phi^+_T$  &$ \Phi^+_S\Psi^-_T$ &$\Phi^+_S\Psi^-_T$ \\ 
        $\Psi^-_T$&         $\Psi^-_T$ & $(-1)\ \ \ \Psi^-_T$   &$-\Psi^+_S\Psi^+_T$ & $\Phi^-_S\Psi^+_T$ \\ 
      \hline\hline
\end{tabular}
\normalsize
\label{tab:deutsch}
\end{table}

The key issue to replacing  BCNOT gates by BiSWAP gates 
is how to convert the relationship 
between CNOT and iSWAP gates into a bilateral form.  
The basic relationship between the CNOT gate 
and the iSWAP gate can be derived by starting from the fundamental 
property that the iSWAP gate can be decomposed into a CNOT gate and 
a SWAP gate between qubits `1' and `2':
\begin{equation}
U_{\rm iswap}=U_{\rm swap}{\rm diag}(1,i,i,1).
\end{equation}
Thus, the relationship between the CPF gate, 
$U_{\rm cpf}={\rm diag}(I,\sigma^z)$ ($I$ is a unit $2\times 2$ matrix) 
and the iSWAP gate can be described as
\begin{equation}
U_{\rm cpf}={\rm diag}(1,i,i,1)P_{1-} P_{2-}
=U_{\rm swap}U_{\rm iswap}P_{1-} P_{2-},
\end{equation}
where $P_{1-}=e^{i\frac{\pi}{4}\sigma^z_1} \otimes I $ and 
$P_{2-}=I \otimes e^{i\frac{\pi}{4}\sigma^z_2}$ 
are $\pi/2$ rotations around the $z$-axis on one of the qubits.
Using $H_1=H \otimes I$ and $H_2=I \otimes H$ 
with the Hadamard matrix  
\begin{equation}
H=\frac{1}{\sqrt{2}}\left(
\begin{array}{cc}
1 & 1 \\
1 & -1 
\end{array}
\right),
\end{equation}
and the relation, 
$ 
U_{\rm cnot}=H_2U_{\rm cpf} H_2
$, 
we have 
\begin{eqnarray}
U_{\rm swap}U_{\rm cnot}&=&U_{\rm swap} H_2 U_{\rm swap}U_{\rm iswap}P_{1-} P_{2-} H_2
\nonumber \\
&=&H_1 U_{\rm iswap} P_{1-} P_{2-} H_2
\label{eq_iswap}.
\end{eqnarray}
We construct a bilateral version of this equation.
The basic strategy is to replace each qubit operation 
by a bilateral one, one by one.
Note that we do not always have to replace each operation 
in Eq.~(\ref{eq_iswap}) by the  
bilateral operation that exactly corresponds to the original unilateral 
operation.
As long as the same effect can be obtained, 
we can instead use a simpler operation.
Then, by observing the roles of each operation, we find the relation:
\begin{equation}
U_{\rm bcnot}=U_{\rm bswap}B^y_{S-} U_{\rm biswap} B^{z}_{S+} B^{z}_{T+} B^{y}_{T+}
\label{eq_biswap}.
\end{equation}
Here, we find that we can replace the Hadamard gates by $B^y_{\pm}$ gates 
by just adjusting the coefficients of the wave functions in the four steps. 
This is the reason why here we introduce $B^x_-$, $B^y_-$ and $B^z_-$, 
in addition to $B^x_+$, $B^y_+$ and $B^z_+$ from Ref.\cite{Bennett}.
Of course, we can express the bilateral Hadamard gate conventionally 
using three single-qubit operations as $B^x_+ B^z_+ B^x_+$. However, 
the Hadamard gate by these three rotations should be avoided so that 
we can reduce the operation time. 
Also note that we do {\it not} need to carry out the SWAP gate in Eq.~(\ref{eq_biswap}), 
because we only have to choose one of the qubits to be measured after 
the BiSWAP gate.
Indeed, the SWAP gate is expressed mathematically by three CNOT gates, 
therefore, faithfully following Eq.~(\ref{eq_biswap}) is against 
our aim of reducing the number of gate operations. 
In the conventional purification process, after the BCNOT gate, 
the `target' qubits are measured and checked whether they are in 
$| \uparrow\uparrow\rangle$ or $| \downarrow\downarrow\rangle$.
In the present case, where we use the iSWAP gate, we measure the 
`source' qubits {\it instead of} the `target' qubits and keep 
the `target' qubits to the next step, if the `source' qubits are  
in $| \uparrow\uparrow\rangle$ or $| \downarrow\downarrow\rangle$.
Here, the SWAP process is  
{\it irrelevant} in this purification process. 
%

The whole pulse sequence in Eq.~(\ref{eq_biswap}) is described 
in Table \ref{tab:deutsch} step by step.  
In step (i), the $B^y_+$ mapping is {\it only} applied to the target qubits.
In step (ii), the $B^z_+$ mapping is applied to {\it both} the source and the target qubits, 
which means that all four qubits are rotated by $\pi/2$ around the z-axis. 
In step (iii), the BiSWAP gate is applied between the source pair and the target 
pair (Fig.1). Finally, in step (vi), the $B^y_-$ mapping is carried out 
on the source pair.
Comparing the right-most column with the expected results of the BCNOT gate, 
we can see that the sequence (\ref{eq_biswap}) is equivalent to the 
BCNOT gate, including its coefficients.
Because we can express $U_{\rm cnot}$ as
\begin{equation} 
U_{\rm cnot}=H_2 P_{1-} P_{2-} U_{\rm iswap}  H_1U_{\rm swap}
\end{equation}
in the reversed order, 
we can also reverse the order of the operation by 
taking the Hermitian conjugate of $U_{\rm bcnot}$ as 
\begin{equation}
U_{\rm bcnot}^\dagger =B^y_{T+}B^z_{S+} B^z_{T+} U_{\rm biswap}  B^y_{S-} U_{\rm bswap}
\label{eq_biswap2}.
\end{equation}
The Deutsch protocol~\cite{Deutsch} 
is more efficient than the Bennett protocol~\cite{Bennett}, because the former 
does not need a Werner state (See below)~\cite{Dur2}.
%
%
In the Deutsch purification protocol, $\pm \pi/2$ rotations around 
the $x$-axis should 
be applied before each BCNOT. Thus, in this protocol, we also have to apply the same 
$\pm \pi/2$ rotations around the $x$-axis before the process shown in Table~\ref{tab:deutsch}.  
If we realize the Deutsch protocol using the conventional CNOT gate Eq.(\ref{time_cnot}), 
then the time needed for each process in the purification protocol is given by :
\begin{equation}
 \tau_{\rm puri}^{\rm bcnot}\approx 5\tau_{\rm rot}+2 \tau_{\rm iswap}.
 \label{puri_bcnot}
\end{equation}
If we replace the CNOT part of the Deutsch protocol by our method, 
we need three single-qubit rotations: $B_{T+}^y$ in step (i), 
$B_{S+}^z$ and $B_{T+}^z$ in step (ii), 
and $B_{T-}^z$ in step (vi), plus an iSWAP gate in step (iii).
From Eq.~(\ref{eq_biswap}) or (\ref{eq_biswap2}), the time $\tau_{\rm puri}^{\rm biswap}$ 
for this entire process in the purification using the BiSWAP is given by
\begin{equation}
 \tau_{\rm puri}^{\rm biswap}\approx 4\tau_{\rm rot}+\tau_{\rm iswap}.
 \label{puri_biswap}
\end{equation}
Thus, the time advantage $\Delta \tau_{\rm puri}^{\rm adv}$ of our method is given by
\begin{equation}
\Delta \tau_{\rm puri}^{\rm adv}=\tau_{\rm puri}^{\rm bcnot}-\tau_{\rm puri}^{\rm biswap}
\approx \tau_{\rm rot}+ \tau_{\rm iswap}
\end{equation}

In the Bennett purification protocol, the mixed-state density matrix 
is assumed to be in a diagonal form called the Werner state:
\begin{equation}
\rho=
 A|\Phi^+\rangle \langle\Phi^+|
+B|\Psi^-\rangle \langle\Psi^-|
+C|\Psi^+\rangle \langle\Psi^+|
+D|\Phi^-\rangle \langle\Phi^-|
\label{eq_bennett}
\end{equation}
with $A=F$ and $B=C=D=(1-F)/3$, where $F$ is the 
fidelity with respect to $\Phi^+$.
The simple form of the density matrix Eq.~(\ref{eq_bennett}) also 
makes our replacement simpler.
This is because, when $B=C=D$, the $B^y_\pm$ mapping does not affect the 
coefficient of the Werner state and, moreover, the diagonal form 
makes the coefficients of the bilateral transformations irrelevant to 
the purification process.
The result is shown in Table~\ref{tab:bennett}. In this case, 
after applying the BiSWAP gate to the initial mixed state 
Eq.~(\ref{eq_bennett}), we can apply either
step (ii-a) involving $B^x_{S\pm}$ $B^x_{T\pm}$ rotations, or 
step (ii-b) involving $B^x_{S\pm}$ $B^x_{T\pm}$ rotations.
Thus, in this case, the protocol only needs two steps.

Here we assume that, for step~(ii-a) in Table~\ref{tab:bennett}, 
the probability of finding $\Phi^+$ is $F$ 
($\Phi^\pm$ can be exchanged into $\Phi^\mp$ by a unilateral $\pi$ 
rotation around the $z$-axis) and 
the probability of finding the other states is $(1-F)/3$.
For step~(ii-b) in Table~\ref{tab:bennett}, the probability of finding the state 
$\Psi^-$ is $F$ ($\Psi^\pm$ can be exchanged by $\Phi^\pm$ by a unilateral 
$\pi$ rotation around the $x$-axis) 
and those of other states are $(1-F)/3$. 
We do not discard the $\Phi^\pm$ elements when measuring the target qubits,
and take $\Phi^-_S$ as the target purified state. 
Then, the probability that the source qubits are in $\Phi^-_S$ after 
this purification process, is 
exactly the same as that by the Bennett protocol~\cite{Bennett}:
\begin{equation}
F'=\frac{F^2+\left(\ \frac{1-F}{3}\right)^2}
{F^2+2F \left(\frac{1-F}{3}\right) +5 \left(\frac{1-F}{3}\right)^2}.
\label{fid_improved}
\end{equation}
The fidelity of the target state is improved ($F'>F$) when $1/2<F<1$.
Thus, we can show that the CNOT gate, which requires {\it two} processes of
qubit-qubit interactions, can be replaced by {\it one} qubit-qubit interaction. 
This is a more efficient purification protocol.

\small

\begin{table}
\caption{Bennett {\it et al}~\cite{Bennett} purification 
process for entangled states. Note that the initial state 
in the first column is subject to the steps shown in the 
following three columns. After applying a BiSWAP in step (i),
the purification process requires applying {\it either} the step (ii-a)
{\it or} the step (ii-b), but not both. 
The ``Test result" columns provide terms, which are used to 
compute fidelities shown in Eq.~(\ref{fid_improved}).
}
\begin{tabular}{r|c|cc|cc}
\hline\hline
initial &  step (i)  & step (ii-a) &    &step (ii-b) & \\ 
 state \  &  BiSWAP  & $B^x_{S\pm}$ $B^x_{T\pm}$ & $\stackrel{\rm Test}{\rm result}$
& $B^y_{S\pm}$ $B^y_{T\pm}$  & $\stackrel{\rm Test}{\rm result}$\\  \hline
$\Phi^+_S\Phi^+_T$& $\Phi^-_S\Phi^-_T$ & $\Phi^-_S\Phi^-_T$& \fbox{$F^2$}                  & $\Psi^+_S\Psi^+_T$& \\ 
        $\Phi^-_T$& $\Phi^+_S\Phi^-_T$ & $\Psi^+_S\Phi^-_T$& $F\left(\frac{1-F}{3}\right)$ &$\Phi^+_S\Psi^+_T$& \\ 
        $\Psi^+_T$&$i\Psi^+_S\Phi^+_T$ &$i\Phi^+_S\Psi^+_T$& &$i\Phi^-_S\Phi^+_T$&\fbox{$\left(\frac{1-F}{3}\right)^2$} \\ 
        $\Psi^-_T$&$i\Psi^-_S\Phi^+_T$ &$i\Psi^-_S\Psi^+_T$& &$i\Psi^-_S\Phi^+_T$&$F\left(\frac{1-F}{3}\right)$\\ 
$\Phi^-_S\Phi^+_T$& $\Phi^-_S\Phi^+_T$ & $\Phi^-_S\Psi^+_T$& & $\Psi^+_S\Phi^+_T$&$\left(\frac{1-F}{3}\right)^2$\\ 
        $\Phi^-_T$& $\Phi^+_S\Phi^+_T$ & $\Psi^+_S\Psi^+_T$& & $\Phi^+_S\Phi^+_T$&$\left(\frac{1-F}{3}\right)^2$\\ 
        $\Psi^+_T$&$i\Psi^+_S\Phi^-_T$ &$i\Phi^+_S\Phi^-_T$& $\left(\frac{1-F}{3}\right)^2$&$i\Phi^-_S\Psi^+_T$&\\ 
        $\Psi^-_T$&$i\Psi^-_S\Phi^-_T$ &$i\Psi^-_S\Phi^-_T$& $\left(\frac{1-F}{3}\right)^2$&$i\Psi^-_S\Psi^+_T$&\\ \hline
$\Psi^+_S\Phi^+_T$&$i\Phi^+_S\Psi^+_T$& $i\Psi^+_S\Phi^+_T$&         $F\left(\frac{1-F}{3}\right)$&$i\Phi^+_S\Phi^-_T$&$\left(\frac{1-F}{3}\right)^2$\\ 
        $\Phi^-_T$&$i\Phi^-_S\Psi^+_T$& $i\Phi^-_S\Phi^+_T$& \fbox{$\left(\frac{1-F}{3}\right)^2$}&$i\Psi^+_S\Phi^-_T$&$\left(\frac{1-F}{3}\right)^2$\\ 
        $\Psi^\pm_T$& $\Psi^\mp_S\Psi^-_T$& discarded& & discarded &\\ 
$\Psi^-_S\Phi^\pm_T$&$i\Phi^\pm_S\Psi^-_T$& discarded& & discarded &\\ 
%
%
        $\Psi^+_T$& $\Psi^-_S\Psi^+_T$ & $\Psi^-_S\Phi^+_T$& $\left(\frac{1-F}{3}\right)^2$ &$\Psi^-_S\Phi^-_T$&$F\left(\frac{1-F}{3}\right)$\\ 
        $\Psi^-_T$& $\Psi^+_S\Psi^+_T$ & $\Phi^+_S\Phi^+_T$& $\left(\frac{1-F}{3}\right)^2$ &$\Phi^-_S\Phi^-_T$&\fbox{$F^2$}\\ 
      \hline\hline
\end{tabular}
\label{tab:bennett}
\end{table}
\normalsize


\subsection{Effect of errors}
Here we check the effect of errors in the Bennett purification process 
shown in Table~\ref{tab:bennett}.
We assume that the XY interaction has a probable pulse error $\epsilon$  
in controlling the interaction time as 
\begin{equation}
2Jt=\pi/2+\epsilon
\end{equation}
($\epsilon \ll 1$) in Eq.~(\ref{iswap}).
Then, for both (a) and (b) columns in Table~\ref{tab:bennett}, we have the relation:
\begin{equation}
F'=\frac{k_1^2 F^2+k_3
\left(\ \frac{1-F}{3}\right)^2}
{k_1 F^2+2F \left(\frac{1-F}{3}\right) +
(5+k_2)\left(\frac{1-F}{3}\right)^2},
\label{fid_improved2}
\end{equation}
where $k_1$, $k_2$ and $k_3$ are given, in second order on the error $\epsilon$ by
\begin{eqnarray}
k_1 &=& (1+\cos 2\epsilon)/2 \ \sim \ (1- \epsilon^2), \nonumber \\
k_2 &=& (1-\cos 2\epsilon)/2 \ \sim \ \epsilon^2, \nonumber \\
k_3 &=& 1+ k_2+ \sin^2(2\epsilon) /4 \ \sim \ (1+2\epsilon^2)
\end{eqnarray}
From these equations, the original condition 
$F> 1/2$ to hold the relation $F'>F$ is changed 
into $F > 1/2+ 3\epsilon^2$, to order $\epsilon^2$. 
Thus, if there is a pulse error,
the initial fidelity for the purification process 
should be correspondingly increased. 

\subsection{Purification using $\sqrt{\rm swap}$ gates}
In the case of the Heisenberg interaction~\cite{Koji}, we cannot directly 
replace the CNOT gate by $\sqrt{\rm SWAP}$
in the purification protocol. This is because $\sqrt{\rm SWAP}$ 
has off-diagonal matrix elements and mixes Bell states.
Thus, for the Deutsch purification protocol, 
we better use the conventional CNOT gate. 
However, for the Bennett protocol, we 
can slightly reduce the number of operations. 
For the Bennett case, we can use the CPF gate plus the $B_{\pm}^y$ 
operation. The CPF gate transforms 
\begin{eqnarray}
\Phi^p \Phi^q &\rightarrow&  \Phi^p    \Phi^q \\
\Psi^p \Psi^q &\rightarrow& -\Psi^{-p} \Phi^{-q} \\
\Phi^p \Psi^q &\rightarrow&  \Phi^{-p} \Psi^q \\
\Psi^p \Phi^q &\rightarrow&  \Psi^{-p} \Phi^{-q},
\end{eqnarray}
where $p=\pm, q=\pm$. 
By combining the CPF gate with $B_{\pm}^y$, we can obtain the same equation 
as Eq.(\ref{fid_improved}) for the $\Phi^+$ state. 
In this case, the advantage is just the time $\tau_{\rm rot}$ to perform 
a single-qubit rotation. 


\section{Replacement of BCNOT by BiSWAP in hashing and breeding protocol} 
\label{sec-hashing}

The replacement of a BCNOT by a BiSWAP 
gate shown in the previous section can also be applied to 
more general cases where the BCNOT gate is used.
Indeed, the BCNOT gate can be automatically 
replaced with the BiSWAP by the following procedure, 
using Eq.~(\ref{eq_biswap}) or Eq.~(\ref{eq_biswap2}).
This replacement process is more transparent and more formal than 
the purification process in the previous section.  
The procedure of replacement is
as follows:\\
(i)~Apply SWAP gate just after each BCNOT gate.\\
(ii)~Replace a BCNOT gate with a BiSWAP gate by Eq.~(\ref{eq_biswap}) 
or Eq.~(\ref{eq_biswap2}),\\
(iii)~Contract a series of $B^x_{\pm}$, $B^y_{\pm}$, $B^z_{\pm}$  
and other single qubit rotations to reduce the number of 
gate operations. \\

In the following two subsections, we apply 
this method to the hashing and the breeding 
protocols proposed by Bennett {\it et al}~\cite{Bennett}.
Note that the process (i) does not mean that 
an additional SWAP gate is needed. That is, we can 
perform the numbering of output qubits without adding 
real gates.

\subsection{Hashing using iSWAP gates}

\begin{figure}
\includegraphics[width=8.6cm]{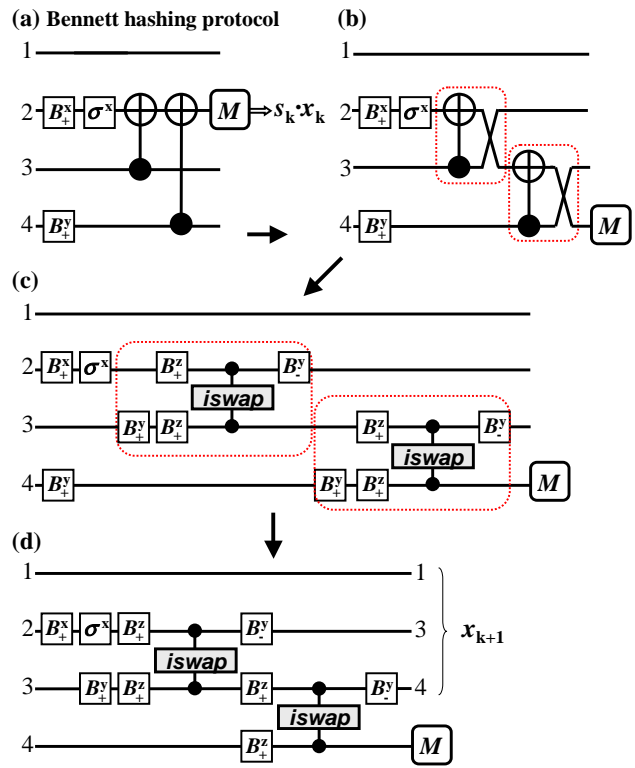}
\caption{Replacement of a BCNOT gate by a BiSWAP gate in the {\it hashing} protocol.
This figure shows the protocol for one of the parties. The complete protocol 
is achieved by the executing the same operation at both ends.
The {\bf {\it`M'}} denotes measurement.
} 
\label{hashing}
\end{figure}

The hashing protocol proposed by Bennett {\it et al}.~\cite{Bennett} is 
based on a one-way communication from Alice 
to Bob (See Fig.~\ref{hashing} (a)).  
In Fig.~\ref{hashing}, $\sigma^x$, $\sigma^y$ and $\sigma^z$ 
express unilateral $\pi$ rotations of one particle.
A sequence of unknown impure pairs, such as 
$\Psi^-\Phi^+\Phi^-...$ is regarded as a bit string 
110010... by the definition,
\begin{equation} 
\Phi^+=00,\ \Psi^+=10,\ \Phi^-=01,\ \Psi^-=11.
\end{equation}
At the $k$-th round of the hashing protocol for 
an initial set of $n$ impure pairs, 
Alice first sends Bob a random $2(n-k)$-bit 
string $s_k \in \{00, 11, 01, 10 \}$ for 
the unknown $(n-k)$ impure pairs $x_k$. 
Depending on the value of $s_k$, gate operations 
for each pair are carried out following Fig.~\ref{hashing}(a).
Then, a parity of this random bit string is 
obtained by the measurement. 
Depending on the parity, 
the probability of the impure pair is reduced 
and we can increase the purity of the resulting states.

When we apply this hashing protocol by a local quantum computer, 
at least four qubits are required for the local quantum computer. 
We can simplify the hashing process after the purification,
by using the BiSWAP gate as follows.
We can replace each BCNOT gate by a BiSWAP gate, one by one, 
as shown in Fig.~\ref{hashing}(b)-(d).
In Fig.~\ref{hashing}(b), we first change the protocol such 
that a SWAP gate follows a CNOT gate. 
In Fig.~\ref{hashing}(c), a CNOT gate plus a SWAP gate is
replaced by an iSWAP gate as shown in Fig.~\ref{replace} and Eq.~(\ref{eq_biswap}).
Because $B^y_+ \Phi^- =-\Psi^+$ and $B^y_+\Psi^+ =\Phi^-$,
we can substantially neglect the effect of $(B^y_+)^2$ in Fig.~\ref{hashing}(d). 

\begin{figure}
\includegraphics[width=8.6cm]{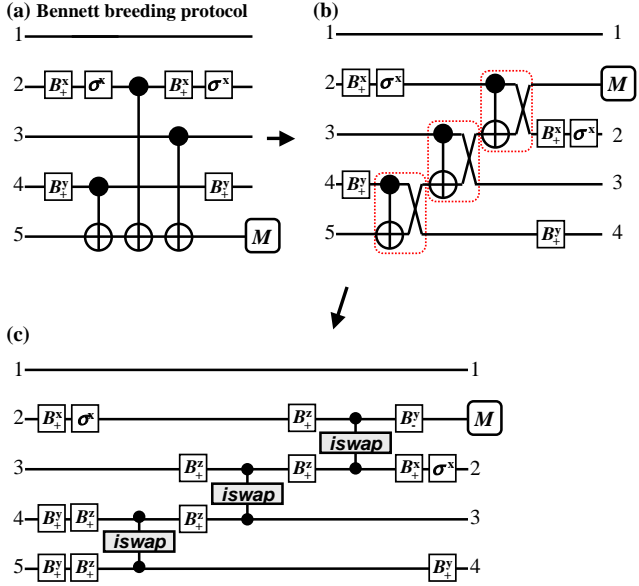}
\caption{Replacement of a BCNOT gate by a BiSWAP gate in the {\it breeding} protocol.
This figure 
shows the protocol for one of the parties. The complete protocol 
is achieved by the executing the same operation at both ends.} 
\label{breed}
\end{figure}

\subsection{Breeding protocol using iSWAP gates}

Here we show an effective way of carrying out the 
breeding protocol proposed by Bennett {\it et al}.~\cite{Bennett} (Fig.~\ref{breed}(a)).
The difference between the breeding and the hashing protocols is 
that, in the former case, Alice and Bob purify 
a sequence of impure states using a pool of initially 
prepared pure states, and the impure pairs do not 
have to be measured. 
Thus, the number of candidates 
of the impure set $x$ is reduced by 1/2 for each breeding 
process, although pure Bell states should be prepared in advance. 
In this breeding protocol,
three CNOT gates are required per each single process. 
Thus, we need three iSWAP gates in order to
replace CNOT gates by iSWAP gates.
Figures~\ref{breed}(b) and (c) show the process of this replacement. 
First, the SWAP gate is inserted just after each 
CNOT gate~(Fig.~\ref{breed}(b)). 
Note that, in Fig.~\ref{breed}(b), we have replaced each original 
two CNOT by a iSWAP gate between nearest qubits. 
Next, each pair of CNOT and SWAP gates
is replaced by a set of iSWAP and single-qubit rotations, 
according to Eq.~(\ref{eq_biswap}). Finally, a series of $B^y$ gates 
are contracted. Then, we obtain 
the breeding protocol using iSWAP gates. 
 
\section{Generation of Bell states} \label{sec-Bell}
In the previous sections, we have assumed that Bell states are initially 
prepared and distributed to two parties. 
Here, assuming a situation that four Bell states should be 
generated by local quantum computers,
we show an effective way of generating the four Bell states 
by an iSWAP gate and $\sqrt{\rm swap}$ gate. 
Conventionally, the Bell states are produced by 
applying the CNOT gate to product states such as
\begin{equation} 
U_{\rm cnot}(|0\rangle_S + |1\rangle_S)|1\rangle_T = |01\rangle +|10\rangle.
\end{equation}
When we use the iSWAP gate, Bell states can be generated by turning-on one iSWAP gate 
with $\pm\pi/2$ rotations around the $y$-axis
\cite{iswap} as follows:
\begin{eqnarray}
e^{i\frac{\pi}{4} \sigma^y_2} U_{\rm iswap}^{(12)} |-\rangle_{y1}|-\rangle_{y2}
&=& |0\rangle_1 |1\rangle_2 -|1\rangle_1 |0\rangle_2 \\ 
e^{-i\frac{\pi}{4} \sigma^y_2} U_{\rm iswap}^{(12)} |-\rangle_{y1}|-\rangle_{y2}
&=& |0\rangle_1 |0\rangle_2 +|1\rangle_1 |1\rangle_2 \\
e^{i\frac{\pi}{4} \sigma^y_2} U_{\rm iswap}^{(12)} |+\rangle_{y1}|-\rangle_{y2}
&=& |0\rangle_1 |1\rangle_2 +|1\rangle_1 |0\rangle_2 \\ 
e^{-i\frac{\pi}{4}\sigma^y_2} U_{\rm iswap}^{(12)} |+\rangle_{y1}|-\rangle_{y2}
&=& |0\rangle_1 |0\rangle_2 -|1\rangle_1 |1\rangle_2 
\end{eqnarray} 
where $|\pm\rangle_y \equiv |0\rangle \pm i|1\rangle$ are 
eigenstates of $\sigma^y$ and 
$U_{\rm iswap} |-\rangle_{y1}|-\rangle_{y1}$ is a two-qubit 
cluster state shown in Ref.~\cite{iswap,Briegel}.
If we start from a product state $|00\rangle$, we need 
two rotations and one iSWAP gate to create four Bell states. In these 
cases, we conventionally need an operation time
\begin{equation}
\tau_{\rm Bell}^{\rm cnot}\approx 
5\tau_{\rm rot}+2\tau_{\rm iswap}.
\end{equation} 
In the present method, we just need :
\begin{equation}
\tau_{\rm Bell}^{\rm iswap}\approx 2 \tau_{\rm rot}+\tau_{\rm iswap}.
\label{eq:new_bell_iswap}
\end{equation}
Therefore, the time advantage is given by
\begin{equation}
\Delta \tau^{\rm adv:iswap}_{\rm Bell}=
\tau_{\rm Bell}^{\rm cnot}-\tau_{\rm Bell}^{\rm iswap}
\approx 3\tau_{\rm rot}+\tau_{\rm iswap}.
\end{equation}
Thus, we can reduce the time $\Delta \tau^{\rm adv:iswap}_{\rm Bell}$ 
for generating the Bell states. 
Similarly, we can produce the Bell states by a single-use of 
$\sqrt{\rm SWAP}$.
Because of the relation
\begin{equation}
U_{\sqrt{\rm swap}}^{(12)} |+\rangle_1|-\rangle_2
= |0\rangle_1 \{|0\rangle_2 + i|1\rangle_2 \} 
-i|1\rangle_1 \{|0\rangle_2 - i|1\rangle_2 \} ,
\end{equation}
if we apply $e^{\mp i\frac{\pi}{4}\sigma^z_1}$
on qubit `1' and 
$e^{i\frac{\pi}{4}\sigma^x_2}e^{i\frac{\pi}{4}\sigma^y_2}$ 
on qubit `2', we obtain $\Psi^{\pm}$. If we apply 
$e^{\mp i\frac{\pi}{4}\sigma^z_1}$ on qubit `1' and 
$e^{-i\frac{\pi}{4}\sigma^x_2}e^{i\frac{\pi}{4}\sigma^y_2}$ 
on qubit `2', we obtain $\Phi^{\pm}$.  
In these cases, we can reduce the time as 
\begin{equation}
\tau_{\rm Bell}^{\sqrt{\rm swap} } \approx 3 \tau_{\rm rot}+\tau_{\sqrt{\rm swap}},
\label{eq:new_bell_rswap}
\end{equation}
compared with the conventionally necessary time
\begin{equation}
\tau_{\rm Bell}^{\rm cnot}\approx 
4\tau_{\rm rot}+2\tau_{\sqrt{\rm swap}}.
\end{equation} 
Thus, the time advantage now becomes
\begin{equation}
\Delta \tau^{{\rm adv:}\sqrt{\rm swap}}_{\rm Bell}=
\tau_{\rm Bell}^{\rm cnot}-\tau_{\rm Bell}^{\sqrt{\rm swap}}
\approx \tau_{\rm rot}+\tau_{\sqrt{\rm swap}}.
\end{equation}
Table~\ref{tab:adv} summarizes operation time advantage discussed in this paper.

\begin{table}
\caption{Summary of the operation time improvements by using our proposed method. 
For $\tau_{\rm puri}^{\rm biswap}$ see Eq.(\ref{puri_biswap}), 
for $\tau_{\rm Bell}^{\rm iswap}$, see Eq.(\ref{eq:new_bell_iswap}), and 
for $\tau_{\rm Bell}^{\sqrt{\rm swap}} $ see Eq.(\ref{eq:new_bell_rswap}).
}
\begin{center}
\begin{tabular}{l|l|l}
\hline\hline
 \ \ \ \ new operation time    &\ \  previous operation time  \ & \ time advantage \\ \hline
 \ $\tau_{\rm puri}^{\rm biswap}\approx 4\tau_{\rm rot}+\tau_{\rm iswap}$\ 
&\ $\tau_{\rm puri}^{\rm bcnot}\approx 5\tau_{\rm rot}+2 \tau_{\rm iswap}$\
&\ \ \ $\tau_{\rm rot}+\tau_{\rm iswap}$\ \\
 \ $\tau_{\rm Bell}^{\rm iswap}\approx 2\tau_{\rm rot}+\tau_{\rm iswap}$ \
&\ $\tau_{\rm Bell}^{\rm cnot} \approx 5\tau_{\rm rot}+2\tau_{\rm iswap}$\  
&\ \ \ $3\tau_{\rm rot}+\tau_{\rm iswap}$ \ \\
 \ $\tau_{\rm Bell}^{\sqrt{\rm swap}} \approx3 \tau_{\rm rot}+\tau_{\sqrt{\rm swap}}$\ 
&\ $\tau_{\rm Bell}^{\rm cnot}\approx 4\tau_{\rm rot}+2\tau_{\sqrt{\rm swap}}$\
&\ \ \ $\tau_{\rm rot}+\tau_{\sqrt{\rm swap}}$\ \\
\hline\hline
\end{tabular}
\end{center}
\label{tab:adv}
\end{table}

\section{Application of the iSWAP purification process} 
\label{sec-example}

In this Section we quantitatively examine several examples using the XY interaction 
and compare our method with the conventional ones in the literature, 
which are based on the CNOT gate.

(1) 
Imamoglu {\it et al.}~\cite{Imamoglu} proposed a quantum computing architecture 
where localized electron spins in QDs are qubits, and they interact with 
each other via the coupling to the vacuum field of a common microcavity. 
In this case, the qubit-qubit interaction mediated by the cavity photon 
is expressed by the XY model with $J=g^2/\Delta$, 
where $\Delta$ is two-photon detuning and 
$g$ is an effective two-photon coupling coefficient for the spin qubits. 
Based on the parameters in the proposal, it takes about 30~psec per each iSWAP gate
and 10~psec per each single-qubit rotation, so that it takes $\sim$100~psec 
for the CPF gate operation.  If we assume that two rotations should be 
added to the CPF gate in order to obtain the CNOT gate, 
it takes about 120~psec for the CNOT operation.  Now if we replace the CNOT gate 
by the iSWAP gate, we need 60 psec in total from 
Eq.~(\ref{puri_biswap}).  Thus the operation time of our method is about half 
of the conventional method. 

(2)
When two superconducting charge qubits 
interact with each other via capacitive coupling to a common superconducting 
coplanar resonator, the resulting effective inter-qubit interaction
is also described by the XY model~\cite{Houck}.  Using $g/\Delta=0.1$, 
$g/(2\pi)=200$~MHz, $\Delta/(2\pi)=2$~GHz, we 
have $J/(2\pi)=20$~MHz and $\tau_{\rm iswap}=6.25$~nsec.  With 
$\omega_{\rm rot}/(2\pi) \sim 1$~GHz, we have $\tau_{\rm rot}\sim 125$~psec. 
Thus, we have $\tau_{\rm puri}^{\rm bcnot}\sim 13.1$~nsec and 
$\tau_{\rm puri}^{\rm biswap}\sim 6.75$~nsec for a dephasing time of about  
500~nsec.  This means that {\it our purification method is about twice 
as fast as the conventional one for entanglement purification}.
We can also apply our method to purify flux qubits connected 
by a common LC circuit data bus~\cite{Liu2}. 

In the cases where multiple qubits are connected by a common cavity field 
or data bus, when we want to purify qubits by the method mentioned above, 
we can only choose one two-qubit pair at a time, 
since we cannot control more than three qubits simultaneously.

\begin{figure}
\includegraphics[width=7cm]{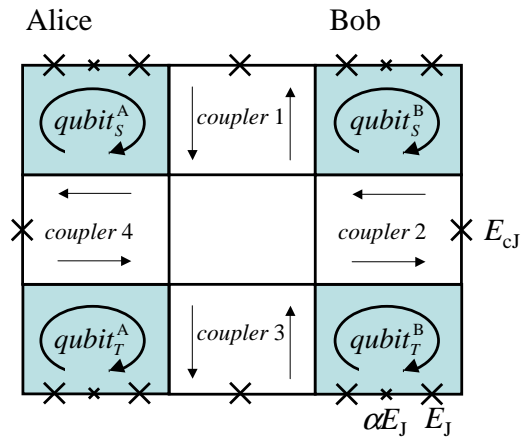}
\caption{Four flux qubits coupled by four couplers. Initially,
the upper two qubits (`source' qubits) and the lower two qubits 
(`target' qubits) are entangled, 
respectively, forming in mixed states. After the purification 
process, the upper `source' qubits are $\sigma^z$-measured.} 
\label{flux}
\end{figure}

(3)
Our method can be applied to purify  
solid-state qubits with
XY interactions, with or without cavity photons. 
Figure~\ref{flux} shows four three-junction superconducting flux-qubits,
coupled to their neighbors via single Josephson junction couplers.
This setup is obtained by extending the setup shown 
in Refs.~\cite{Grajcar,Niskanen,Ashhab,Liu3,Yamamoto3}. 
We take $E_{\rm cJ} > E_J$ and $0.5 < \alpha < 1$ 
such that only the ground state of the four couplers 
(classical region) is involved in the coupling process,
and each of the three-Josephson-junction loops constitutes a flux qubit.  
Here we can consider the purification process of an entangled state 
between qubit${}_{T}^A$ and 
qubit${}_{T}^B$ using qubit${}_{S}^A$ and qubit${}_{S}^B$, by controlling 
the four couplers that exist between each pair of neighboring qubits.
If we use experimental values $J/(2\pi) \sim 25$~MHz~\cite{Grajcar} and 
a single qubit frequency of 1~GHz~\cite{Izmalkov}, the gate times are  
$\tau_{\rm iswap}\sim 5$~nsec and $\tau_{\rm rot}\sim 125$~psec.
%
Thus, $\tau_{\rm puri}^{\rm bcnot} \sim 10.6$~nsec and 
$\tau_{\rm puri}^{\rm biswap} \sim 5.5$~nsec, for a qubit dephasing time of 
500~nsec. 
Depending on the measurement time, we can probably carry out more than one 
purification process well within the qubit coherent time.
After the purification process, we measure the `source' qubits.
If the measured results are in the $|\uparrow\uparrow\rangle$ 
or $|\downarrow\downarrow\rangle$ states, we 
can expect that the entangled state has been improved.  
Otherwise, we restart the whole process by again preparing  
mixed states for the two pairs.

(4) 
For charge qubits based on capacitively-coupled single-electron QDs, 
the inter-qubit XY interaction appears in a rotating reference frame 
when an oscillating gate bias is applied~\cite{tana01}.
For coupled QDs where the radius of each QD is about 2.5~nm and the distance 
between qubits is about 12~nm, $J\sim 0.1$~meV and $\omega_{\rm rot}\sim 0.8$~meV. 
If we assume that we can switch on and off the coupling between QDs, 
we have $\tau_{\rm puri}^{\rm bcnot}\sim 85.3$~psec and 
$\tau_{\rm puri}^{\rm biswap}\sim 48.7$~psec for a dephasing time of 
 100 nsec~\cite{Wiel}. 


\section{Discussion} 
\label{sec-discussion}

We have shown how to effectively reduce the number of operational steps 
in the purification protocol. In any stage of quantum communication, 
all efforts to speed up each process are strongly recommended
from the viewpoint of finite coherence time as well as 
user's satisfaction. The recent cavity-QED techniques 
using superconducting circuits have realized the strong coupling 
between the cavity-mode and the qubit~\cite{You,Houck}. 
The present method of reducing the number of operations is effective 
for all qubits with XY interaction and 
would be of great use to realize quantum communication.

One of the possible quantum communication systems contains local 
quantum computers
based on cavity-QED mechanism and an optical fiber using photons.
This is because the optical fibers would be the lowest cost and most 
effective medium between distant parties, 
and the cavity-QED mechanism is effective to connect 
photon to a local electronic system~\cite{Cirac}. 
Thus, the effective transformation between local quantum states and 
photons is desirable. 
Houck {\it et al.}~\cite{Houck} have succeeded in controlling microwave 
photons in a superconducting circuit based on charge qubits. 
On the other hand, QDs are also good resources of entangled 
photon states~\cite{Andrew}. 
More experiments regarding the emission and absorption 
of photons between the local cavity-QED system and the external 
photonic system are desired. 

In section~\ref{sec-example}, we have shown four examples of applying 
the proposed purification protocols to solid state qubits. The bottom line of 
using the purification protocol is whether we can prepare mixed states 
in which the probability of the desired Bell state is more than 1/2. 
At the first stage of quantum communication, we try to generate 
desired entangled states. 
However, those states are mostly imperfect and decohere gradually. 
If the probabilities of those entangled states are more than 1/2 
even after passing through noisy channels, we can apply the 
purification protocol on those impure pairs. 
In order to repeat the next purification process, 
the time $\tau_{\rm puri}^{\rm biswap}$  $+\tau_{\rm meas}$ 
should be sufficiently smaller than the coherence time 
( $\tau_{\rm meas}$ is a measurement time for judging the two-qubit 
states, which is, for example, 1$\sim$10~msec in Ref.~\cite{Izmalkov}). 
Otherwise, it is possible that the revised fidelity by the purification
is smaller than that of the original state.
In the third example in section~\ref{sec-example}, 
for the second purification process, 
we have to generate a new mixed state from the measured qubits 
(called `source' qubits).  
The measured qubits are in a product state, $|\uparrow\downarrow\rangle$ or 
$|\downarrow\uparrow\rangle$.
First, we try to make a desired entangled state 
using the method mentioned in section~\ref{sec-Bell}.
If the noisy environment successfully changes 
the imperfect entangled state into a mixed 
state with $A>1/2$ in Eq.(\ref{eq_bennett}), 
we can proceed to the next purification. 
Otherwise, we have to apply random 
$B^x$, $B^y$ and $B^z$ rotations. 
Because it takes a time $\tau_{\rm rot}$ 
for each rotation, the total time 
to carry out the second purification process 
is given by 
$\tau_{\rm Bell}^{\rm iswap}+n_{\rm rot}\tau_{\rm rot}$ ($n_{\rm rot}\ge 0$ is 
an integer for the randomization ).
This time should be smaller than that of the coherence 
time of the other surviving qubit~(called `target' qubit) that is waiting 
for the new entangled qubit. 
Whether these purification protocols succeed or not seems to strongly 
depend on each decoherence mechanism.

In this paper, we did not include any quantum error-correcting code 
in the purification protocol. This is because the quantum error-correcting code 
requires many qubits, in contrast to the current experimental situation 
with very few solid-state qubits.   
How to effectively combine the proposed purification process 
with the various quantum error-correcting codes 
would be an important issue for future studies.
%
%

\section{Summary} 
\label{sec-summary}

In summary, we have constructed an efficient adaptation 
of the entanglement purification protocols 
for qubits with XY interactions. 
Specifically, we show that the conventional CNOT gate, which requires 
turning on {\it two}-qubit interactions {\it twice}, can be replaced by a 
{\it single} iSWAP gate together with single-qubit rotations.  This
simplification of the gate pulse sequence reduces the time for
entanglement purification and increases the robustness of the
protocols.
Our method could be used for any qubits 
with XY interactions, particularly
cavity-coupled qubits, which allows solid-state qubits to be more 
easily integrable into a quantum communication network.

\acknowledgements
FN and XH are supported in part by the US
National Security Agency, 
Laboratory for Physical Sciences, 
Army Research Office, and the 
National Science Foundation.
We thank A. Nishiyama, S. Fujita and S. Ishizaka for useful discussions.

\appendix

\section{XY model}
Here, we summarize the derivation of the XY interaction between 
qubits in a cavity~\cite{You,Liu2}. 
The Hamiltonian of two qubits in cavity is given by 
the Jaynes-Cummings Hamiltonian:
\begin{equation}
H_{\rm JC}=\omega a^\dagger a + \sum_{i=1}^{2} \left\{ \frac{\omega_{qi}}{2} \sigma^z_i 
+ (\chi_i \sigma^+_i a + {\rm H.c.}) \right\},
\end{equation}
where the qubit operators are defined by 
$\sigma^z_i=|e\rangle_i \langle e|_i-|g\rangle_i \langle g|_i$, 
$\sigma^+_i=|e\rangle_i \langle g|_i$ and $\sigma^-_i=|g\rangle_i \langle e|_i$ 
using its ground $|g\rangle_i$ and first excited $|e\rangle_i$ states.
In order to derive the two-qubit interaction, a unitary transformation 
$U=\exp(S)$ 
with 
\begin{equation}
S=\sum_{i=1,2} \alpha_i (a^\dagger \sigma^-_i -a \sigma^+_i )
\end{equation}
is introduced.  
For small parameters $\alpha_1$ and $\alpha_2$, the Hamiltonian 
is transformed in second order in $S$ such that
\begin{equation}
H'_{\rm JC}=e^S H_{\rm JC} e^{-S}\approx H_{\rm JC}+[S,H_{\rm JC}]+\frac{1}{2}
[S,[S, H_{\rm JC}]].
\end{equation}
The value of $\alpha_i$ $(i=1,2)$ is determined such that the linear coupling terms 
between $a$ and $\sigma^{\pm}$ are deleted and $\alpha_i=\chi_i/\Delta_i$
with 
\begin{equation}
\Delta_i=\omega-\omega_{qi}.
\end{equation}
Then, we have
\begin{equation}
H'_{\rm JC} \approx  \omega a^\dagger a 
+\sum_{i=1}^{2} \frac{\tilde{\omega}_q}{2}\sigma_i^z  \nonumber \\
+\frac{\chi_1\chi_2(\Delta_1+\Delta_2)}{2\Delta_1\Delta_2}
( \sigma_1^+\sigma_2^- + \sigma_1^-\sigma_2^+)
\end{equation}
with 
\begin{equation}
\tilde{\omega}_{qi}=\omega_{qi}+\chi_i^2/\Delta_i.
\end{equation}
Thus we obtain the XY model from the Jaynes-Cummings model 
with a interaction strength of 
\begin{equation}
J=[\chi_1\chi_2(\Delta_1+\Delta_2)]/(4\Delta_1\Delta_2).
\end{equation}

\end{document}